\documentclass{elsart}

\usepackage{amssymb}
\usepackage{amsmath}
\usepackage[dvips]{graphicx}
\usepackage{bm}

\newcommand{\be}{\begin{equation}}
\newcommand{\ee}{\end{equation}}

\begin{document}

\begin{frontmatter}

\title{Fast Rydberg gates without dipole blockade\\ via quantum control}

\author[TN]{M.~Cozzini},
\author[TN,Harvard]{T.~Calarco},
\author[TN]{A.~Recati},
\author[Innsbruck]{P.~Zoller}

\address[TN]{Dip. di Fisica, Universit\`a di Trento and BEC-CNR-INFM,
I-38050 Povo, Italy}
\address[Harvard]{ITAMP, Harvard-Smithsonian Center for Astrophysics,
Cambridge,\\ MA 02138, USA}
\address[Innsbruck]{Institute for Quantum Optics and Quantum Information of the Austrian
Academy of Sciences
and Institut f\"ur Theoretische Physik, Universit\"at Innsbruck, A-6020
Innsbruck, Austria}

\begin{abstract}

We propose a scheme for controlling interactions between
Rydberg-excited neutral atoms in order to perform a fast
high-fidelity quantum gate. Unlike dipole-blockade mechanisms
already found in the literature, we drive resonantly the atoms
with a state-dependent excitation to Rydberg levels, and we
exploit the resulting dipole-dipole interaction to induce a
controlled atomic motion in the trap, in a similar way as
discussed in recent ion-trap quantum computing proposals. This
leads atoms to gain the required gate phase, which turns out to be
a combination of a dynamic and a geometrical contribution. The
fidelity of this scheme is studied including small anharmonicity
and temperature effects, with promising results for reasonably
achievable experimental parameters.

\end{abstract}


\maketitle
\end{frontmatter}


\section{Introduction}
\label{sec:intro}

The quest for the reliable implementation of a fundamental quantum
gate is one of the central topics in the current research in
quantum computation. The basic requirement is the controlled
coherent dynamics \cite{Shore} of the system realizing the quantum
bits, while avoiding decoherence due to fluctuations of the
external control parameters and coupling to an
environment%
\footnote{We dedicate this publication to Bruce Shore as one of the pioneers in
developing the theory of coherent interactions between atoms and laser light.}.

In the last years, several proposals for different physical
systems have appeared, ranging from trapped ions \cite{ions},
photons in cavity QED \cite{QED}, molecules \cite{molecules}, to
quantum dots and Josephson Junctions \cite{qdots and JJ}. A
particularly attractive perspective is the possibility of using
cold neutral atoms \cite{neutral}, whose steadily improving
experimental control looks quite promising. Regarding the required
logical operations, different possible universal sets of quantum
gates have been devised, showing that single-qubit and two-qubit
gates are sufficient to realize any $N$-qubit unitary operator. In
this context, a widely studied fundamental gate is the two-qubit
controlled-phase gate, whose truth table is given by
\begin{eqnarray}
|g\rangle|g\rangle & \to & \mathrm{e}^{\mathrm{i}\phi}|g\rangle|g\rangle \nonumber \\
|g\rangle|e\rangle & \to & |g\rangle|e\rangle \nonumber \\
|e\rangle|g\rangle & \to & |e\rangle|g\rangle \nonumber \\
|e\rangle|e\rangle & \to & |e\rangle|e\rangle \ ,
\label{eq:truth}
\end{eqnarray}
where $\phi$ is the phase and $|g\rangle$, $|e\rangle$ are the
logical states. When $\phi$ equals $\pi$, this gate is equivalent
to a Controlled-NOT and is universal if combined with arbitrary
single-qubit rotations.

An important figure of merit to evaluate a specific physical
implementation is the ratio between the gate operation time and
the coherence time characteristic of the system. In this sense,
for a given quantum memory coherence time and gate error rate,
faster gates are of course highly preferable. However, gate speed
is limited by several system-specific factors, for instance by the
characteristic energy scales that are present: roughly speaking,
if the phase $\phi$ in Eq.~(\ref{eq:truth}) is produced by a
state-dependent energy shift $\Delta E$, the time it takes for a
phase $\pi$ to be accumulated has to be of the order of $\Delta
E^{-1}$.

In trapped atomic systems the relevant energy scales are the
frequency of the trapping itself (which can hardly exceed, say, a
few hundred kHz) and the energy of atom-atom interactions. This
has led to several theoretical explorations towards employing
stronger interactions like dipole-dipole forces between
Rydberg-excited atoms \cite{jaksch} or molecular interactions
giving rise to Feshbach resonances \cite{Feshbach}. The main issue
to be dealt with in this case is that strong interactions tend to
perturb significantly the trapped-atom dynamics, thereby
representing a potential source of errors. Ways around this
problem depend again on the specific feature of the system under
consideration. For instance, putting Rydberg atoms in a
sufficiently stable and intense static electric field, a
dipole-blockade mechanism can arise that shifts the two-atom
interacting state out of resonance from single-atom exciting
lasers \cite{weidemueller,gould}. In this case, a two-qubit gate
can be performed without ever switching on the dipole-dipole
interaction, thereby avoiding decoherence due to entanglement of
the qubit with the motion.

An alternative way to overcome trap-limited gate time scales is
to employ quantum control techniques, as originally proposed in
the context of ion-trap quantum gates in \cite{garcia}. In this
approach, the gate phase is not directly or indirectly obtained
from a state-selective energy shift due to interaction, but rather
by driving the two-particle system on a closed path in phase
space, whose final result is to bring it back to its original
state, while at the same time imprinting on it the desired phase.

The present paper represents an application of these methods,
developed in \cite{garcia,garcia long} for trapped ions, to the
case of Rydberg-excited atoms already discussed in \cite{jaksch}.
The use of quantum control techniques allows for exploiting the
strong perturbation due to dipole-dipole interaction as a means
for driving the system into the goal state  with a high fidelity
that does not depend, to a good approximation, on the initial
motional state of both atoms inside the trap, and in a time that
is a fraction of the trap period. Moreover, since the phase is not
accumulated as a result of some interaction-induced energy shift
as discussed above, the atoms are not required to remain in the
interacting state for the whole duration of the gate, but just
during much shorter times as needed for imparting the ``kicks''
which drive their motion. In this way  the effect of decoherence
due to spontaneous emission from the highly excited Rydberg
states is further reduced.

A generic quantum computation requires in principle the ability to
perform more complex sequences of two-qubit gates between any pair
of qubits. To achieve this, one will need to move atoms between
different locations in the quantum register. Thus the overall
speed of a complete quantum computation in general will be
basically limited by the time needed to transport the atoms close
to one another before performing each actual gate operation on the
desired qubit pairs. This limitation affects all quantum computing
implementations based on atomic (or ionic) systems. However, it
can be partially overcome by applying quantum control techniques
to the transport process \cite{schmidt-kaler}.
In addition, individual addressing, which is still an issue in present
experiments, is not needed at the level of a two-qubit gate within the protocol
proposed here, and therefore atoms can be kept during the gate at a shorter
distance than required in other schemes by single-atom laser
diffraction-limited addressability.
Thus atom transport would only be required for two-qubit gates involving
non-neighboring qubits, and it may be employed as well whenever a gate on a
single qubit is needed during the computation, to single it out by moving it
away from the other qubits.

The structure of the paper is as follows. In Sec.~II we introduce
in general terms the simple control procedure we plan to apply to
the system for this purpose; in Sec.~III we describe the physical
model for the implementation of the control pulses with
dipole-interacting atomic systems and discuss experimental parameters;
and in Sec.~IV we analyze
quantitatively the obtainable fidelity, before drawing our
conclusions in Sec.~V.

\section{Control scheme}
\label{sec:scheme}

We begin by discussing a possible `kick' scheme to manipulate the system phase
while bringing the atoms back to the initial motional state, in the spirit of
Ref.~\cite{garcia}. Although in the next sections we will adapt this protocol
to our particular configuration, we point out that this scheme is applicable to
quite general systems. In this section, the `system' is simply given by a
single particle in a one-dimensional harmonic potential, corresponding in the
following to the relative motion of the two atoms.

\begin{figure}
\centering
\includegraphics[width=8.5cm]{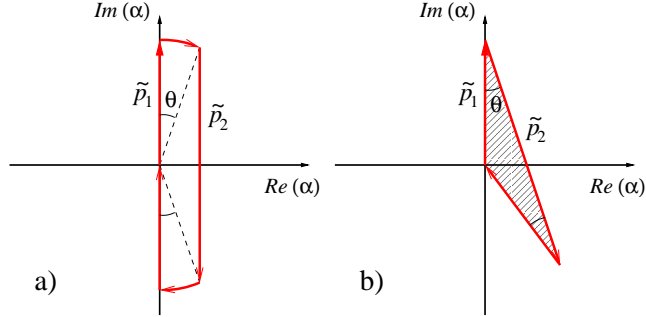}
\caption{\label{fig:triangle}%
Phase space trajectory of a simple kick protocol: a) usual reference frame; b)
rotating reference frame. The harmonic oscillator ground state is first given a
positive momentum $p_1=\sqrt2\tilde{p}_1$; then it is left evolve in the pure
harmonic oscillator potential for a time $\theta$ (in harmonic oscillator
units), and hence kicked again by a negative momentum
$p_2=\sqrt2\tilde{p}_2=-2p_1\cos(\theta)$. The purely harmonic evolution time
$\theta$ corresponds to the angle shown in the figure. The final kick, which
brings the system back to the initial coherent state, has the same intensity
$p_1$ as the first one.}
\end{figure}

The basic idea is to lead the system along a closed path in the oscillator
phase space by taking advantage of the momentum shifts -- the so-called kicks
-- induced by some external force. The simplest possible protocol, i.e., that
requiring the minimal number of kicks, is illustrated in
Fig.~\ref{fig:triangle} for the special case of the oscillator ground state.
This scheme, as far as the kicks are ideal shifts (see the impulse
approximation discussed in the next section), has the remarkable advantage of
being independent of the initial state.

We consider the unitary phase space displacement operator
$D(\alpha)=\mathrm{e}^{\alpha\hat{a}^\dagger-\alpha^*\hat{a}}$, where
$\hat{a}^\dagger$ and $\hat{a}$ are respectively the creation and annihilation
operators of the harmonic oscillator and dimensionless units are used%
\footnote{We recall that $D(\alpha)|0\rangle=|\alpha\rangle$, where $|0\rangle$
is the oscillator ground state and the coherent state $|\alpha\rangle$ is an
eigenstate of the annihilation operator
$\hat{a}|\alpha\rangle=\alpha|\alpha\rangle$.}.
With this formalism, a kick of momentum $p$ is simply given by
$\mathrm{e}^{\mathrm{i}p\hat{x}}=D(\mathrm{i}\tilde{p})$, where $\tilde{p}=p/\sqrt2$ is a real
number and $\hat{x}=(\hat{a}+\hat{a}^\dagger)/\sqrt2$ is the position operator.
The protocol shown in Fig.~\ref{fig:triangle} is a sequence of three kicks
alternated by free harmonic evolution, the latter being governed by the
harmonic time evolution operator
$U_{\mathrm{ho}}(t)=\mathrm{e}^{-\mathrm{i}(\hat{a}^\dagger\hat{a}+1/2)t}$.
By using the well known relations
$D(\alpha)D(\beta)=D(\alpha+\beta)\mathrm{e}^{\mathrm{i}\text{Im}(\alpha\beta^*)}$ and
$U_{\mathrm{ho}}(t)D(\alpha)=D(\mathrm{e}^{-\mathrm{i}t}\alpha)U_{\mathrm{ho}}(t)$, one can
easily rewrite the kick operator
$G=D(\mathrm{i}\tilde{p}_3)U_{\mathrm{ho}}(\theta_2)D(-\mathrm{i}\tilde{p}_2)
U_{\mathrm{ho}}(\theta_1)D(\mathrm{i}\tilde{p}_1)$
in the form
\be
G=U_{\mathrm{ho}}(\theta_1+\theta_2)
D(\mathrm{i}[\tilde{p}_1-\tilde{p}_2\mathrm{e}^{\mathrm{i}\theta_1}+\tilde{p}_3
\mathrm{e}^{\mathrm{i}(\theta_1+\theta_2)}])\mathrm{e}^{\mathrm{i}\phi_{\mathrm{geom}}} \ ,
\ee
where
$\phi_{\mathrm{geom}}=\tilde{p}_1\tilde{p}_3\sin(\theta_1+\theta_2)-
\tilde{p}_1\tilde{p}_2\sin\theta_1-\tilde{p}_2\tilde{p}_3\sin\theta_2$.
The `triangle' scheme is nothing but the condition
$\tilde{p}_1-\tilde{p}_2\mathrm{e}^{\mathrm{i}\theta_1}+\tilde{p}_3
\mathrm{e}^{\mathrm{i}(\theta_1+\theta_2)}=0$,
so that $G=U_{\mathrm{ho}}(\theta_1+\theta_2)\mathrm{e}^{\mathrm{i}\phi_{\mathrm{geom}}}$ and
the initial motional state is restored.
The triangular path is evident in the rotating frame description (see
Fig.~\ref{fig:triangle}(b)), namely the interaction picture where the action of
the oscillator Hamiltonian $H_{\mathrm{ho}}=\hat{a}^\dagger\hat{a}+1/2$ is made
implicit and the gate operator does not contain the term
$U_{\mathrm{ho}}(\theta_1+\theta_2)$.
In addition, one immediately finds
$\tilde{p}_1\tilde{p}_2\sin\theta_1=\tilde{p}_2\tilde{p}_3\sin\theta_2=
\tilde{p}_1\tilde{p}_3\sin(\theta_1+\theta_2)=2A$,
where $A$ is the area of the triangle described by the rotating phase space
trajectory.
Finally, $\phi_{\text{geom}}=-2A$ and, for the considered case
$0\leq\theta_1=\theta_2=\theta\leq\pi/2$,
$\phi_{\text{geom}}=-\tilde{p}_1^2\sin(2\theta)$.
Being based on operatorial identities this
result is evidently independent of the initial state.
It is worth recalling the `geometric' nature of such phase, as discussed in
Ref.~\cite{garcia long}, reflected by the non-commutativity of the phase-space
displacement operators.

We stress that the discussion in this section simply refers to the
dynamics of a 1D harmonic oscillator. Below we will see that such
scheme can be applied to the effective 1D oscillator obtained for
the relative motion of the two considered atoms, so that this
geometric phase will be used in a protocol for a two-qubit gate.
The concrete implementation will however introduce some approximations
whose validity will actually depend on the initial oscillator state (see
Subsec.~\ref{subsec:approximations}).

\section{Physical implementation}
\label{sec:rydberg}

We now turn to the analysis of the particular case where the kicks
are provided by the switching of the dipole-dipole interaction. We
will first give an overview of the actual form of the interaction, then we
will discuss its application to the protocol presented in the
previous section, and finally we will examine the validity
conditions for the approximations introduced.

\subsection{The dipole-dipole interaction}
\label{subsec:dip}

We consider two atoms in two different one-dimensional harmonic
wells with the same frequency $\omega$, separated by a distance
$x_0$. This system can be thought of as an approximate description
of two atoms sitting in different sites of a optical lattice or
in distinct trap minima on an atom chip, where the
dynamics transverse to the direction joining the atoms are completely frozen.
We will come back on such an approximation in
Subsec.~\ref{subsec:approximations}. Each atom corresponds to one qubit, where
two internal hyperfine atomic levels, which we call $|g\rangle$ and
$|e\rangle$, represent the logical states.

\begin{figure}
\centering
\includegraphics[width=8.5cm]{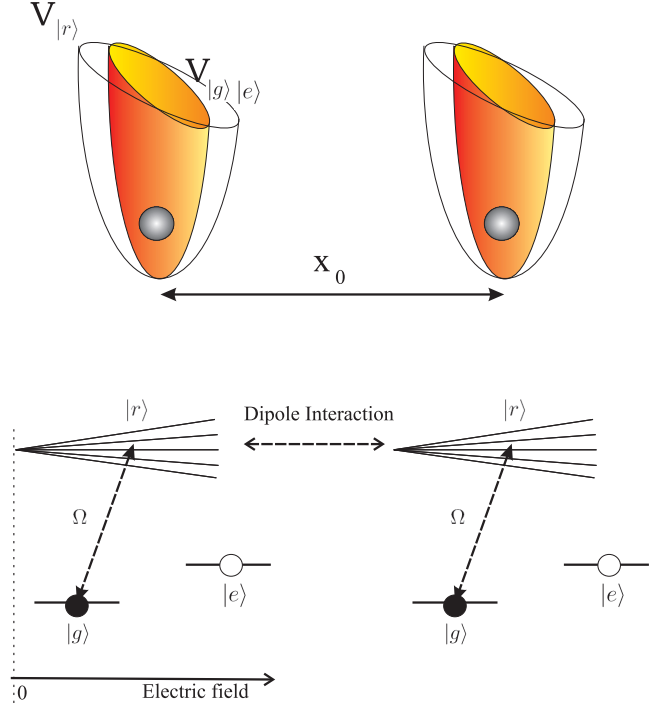}
\caption{\label{fig:levels}
Level scheme for the quantum phase gate. The internal state $|g\rangle$
is coupled by the Rabi frequency $\Omega$ to the Rydberg
state $|r\rangle$. Since the state $|e\rangle$
is unaffected by the laser, the dipole interaction arises only if both
atoms are initially in the $|g\rangle$ state.
The harmonic trapping potential $V_{|r\rangle}$ felt by the atoms in the
Rydberg level $|r\rangle$ is different from the potential
$V_{|g\rangle,|e\rangle}$ of the hyperfine levels $|e\rangle$ and $|g\rangle$.}
\end{figure}

Assuming an underlying static electric field, the gate is
performed by driving state-dependent transitions to Rydberg levels
(see Fig.~\ref{fig:levels}), switching on and off a dipole
atom-atom interaction which allows to manipulate the global phase
of the system, as described in Ref.~\cite{jaksch}. Initially, each
atom can be either in $|g\rangle$ or $|e\rangle$. The internal state
$|g\rangle$ is coupled by a laser to a Rydberg state $|r\rangle$,
where the transition $|g\rangle\to|r\rangle$ is assumed to be
obtained via a Doppler-free two-photon absorption, which
suppresses any photon kick contribution and gives rise to an
effective Rabi frequency $\Omega$ (see Fig.~\ref{fig:levels}). The
state $|e\rangle$ is instead fully decoupled from the Rydberg
manifold. In this way, we switch on a dipole-dipole interaction
only when both atoms are initially in the internal state $|g\rangle$.
This interaction is kept on for a very short time, after which the
atoms are brought back to the $|g\rangle$ states.
While in the $|rr\rangle$ state, the atoms acquire a dynamical
phase as well as a momentum shift (kick).
Thus, repeating this procedure according to the scheme described in Sec. II,
such a process realizes the phase gate (\ref{eq:truth}), provided we undo the
$\pi$-phase due to the Rabi flopping for the states $|ge\rangle$ and
$|eg\rangle$.

The Hamiltonian of the system without the laser coupling between the states
$|g\rangle$ and $|r\rangle$ reads
\begin{eqnarray}
H &=& \sum_{i=1,2}\left[\frac{p_i^2}{2m}+\frac{1}{2}m\omega^2(x_i-x_{0i})^2\right]
(|g\rangle_i\langle g |+|e\rangle_i\langle e |) \nonumber \\
&+& \sum_{i=1,2}\left[\frac{p_i^2}{2m}+\frac{1}{2}m\omega^{\prime 2}(x_i-x_{0i})^2\right]
|r\rangle_i\langle r | \label{eq:H}\\
&+& V_{\rm dip}(x_1-x_2)|r\rangle_1\langle r |\otimes|r\rangle_2\langle r |\nonumber
\end{eqnarray}
where we consider that the (harmonic) trapping potential is the same for the
two hyperfine levels (frequency $\omega$) and different for the Rydberg state
(frequency $\omega'$).
The general 3D expression for the dipole-dipole interaction is
\be
V_{\text{dip}}(\bm{r}) = \frac{1}{4\pi\epsilon_0}
\left[\frac{\bm{\mu}_1\cdot\bm{\mu}_2}{|\bm{r}|^3}-
3\frac{(\bm{\mu}_1\cdot\bm{r})(\bm{\mu}_2\cdot\bm{r})}{|\bm{r}|^5}\right] \ ,
\ee
where $\bm{\mu}_1$, $\bm{\mu}_2$ are the dipole moments of the two atoms,
which depend on the Rydberg level and the external electric field,
and the 3D relative distance $\bm{r}$ reduces in our case to the 1D relative
motion coordinate $x=x_2-x_1$.
In the Hamiltonian (\ref{eq:H}) $m$ is the mass of the atoms and we take
$x_{02}-x_{01}=x_0$, where $x_0$ is the distance separating the two harmonic
wells (see Fig.~\ref{fig:levels}).

The possibility of exploiting this interaction to gain a
non-trivial phase shift and effect a quantum gate has been
suggested in Model A of Ref.~\cite{jaksch}. In this scenario,
during the interaction time $\Delta{t}$, the system acquires a
dynamic phase $u\Delta{t}$, where $u$ is the energy shift created
by the dipole potential. However, in order to reproduce the truth
table (\ref{eq:truth}), one also has to ensure the restoring of
the initial motional states, which are strongly affected by the
non-negligible mechanical effects caused by the dipole interaction
-- and this is not straightforward in this scheme. To overcome
this, in Ref.~\cite{jaksch} a Model B has been introduced, where
the strong dipole-dipole interaction is instead used to induce a
dipole-blockade mechanism that still ensures state-dependent
dynamics while actually avoiding mechanical back-action on the
atomic motion.

An alternative solution can be achieved by controlling such strong
mechanical effects to gain an additional geometrical phase, while
at the same time bringing the system back to the initial motional
configuration, similarly to the method proposed in
Ref.~\cite{garcia}.

In the following we will consider only two possible configurations
for $V_{\text{dip}}(x)$:
the repulsive case $V_+(x)=\mu_1\mu_2/4\pi\epsilon_0x^3$, corresponding to
dipoles aligned in the direction orthogonal to the $x$-axis, and the attractive
case $V_-(x)=-2\mu_1\mu_2/4\pi\epsilon_0x^3=-2V_+(x)$, corresponding to dipoles
parallel to the $x$-axis. This in principle could be achieved by rapidly
changing the electric field orientation between subsequent laser pulses --
although from a practical point of view it may be problematic to realize this
on the time scales required here. A viable alternative would be to keep the
external field fixed and excite the system to Rydberg states having differently
oriented dipole moments, something that could be performed without requiring
individual laser addressing of the atoms.

The expression of the dipole moment for Rydberg states is given by
$\mu_{1,2}=(3/2)n_{1,2}q_{1,2}ea_0$ \cite{gallagher}, where $e$ is the electron
charge, $a_0$ is the Bohr radius, and $n,q$ are quantum numbers which label the
different states.
The quantum number $q$ obeys the relation $q=n-1-|m|,n-3-|m|,...,-(n-1-|m|)$,
where $m$ is the magnetic quantum number. Typical values considered here
(for $^{87}$Rb) are $n_{1,2}=q_{1,2}+1=50$.

We prefer to work with dimensionless quantities by using the oscillator
units of the harmonic potential with frequency $\omega$, felt by the atoms for
most of the time.
The dipole potential in particular can be rewritten as
$V_+/\hbar\omega=\alpha_+^2/(x/a_{\text{ho}})^3$, where
$\alpha_+^2=(9/4)n_1q_1n_2q_2(\mu/m_e)/(a_{\text{ho}}/a_0)$, $\mu=m/2$ is the
reduced mass, $m_e$ is the electron mass, and
$a_{\text{ho}}=\sqrt{\hbar/\mu\omega}$ is the harmonic oscillator length and
then we put $\hbar=\mu=a_{\text{ho}}=1$.

Note that in order to realize an ideal transition to the Rydberg state one
needs the condition $\Omega>>|V_{\pm}(x_0)|$, i.e.,
$\Omega\gg2\alpha_+^2/x_0^3$, which is easily satisfied in all the cases of
practical interest. Realistic values for the parameters are discussed in
Subsec.~\ref{subsec:approximations}.

We describe the action of the dipole potential arising when both atoms are in a
Rydberg state by using an impulse approximation.
We indeed assume that the dipole interaction is very strong compared to the
harmonic trapping and that its duration is very short, so that the system
undergoes a small displacement during this time and the corresponding harmonic
potential contribution to the phase is negligible.
Furthermore we separate the centre of mass and relative motion, where only the
latter is affected by the dipole-dipole interaction.
Due to the very short time spent by the atoms in the Rydberg state,
we then consider the centre of mass motion to be completely unperturbed by the
gate and will only examine the relative motion.
For a short time interval $\Delta{t}$ the relative motion time evolution
operator can hence be approximated by
\be \label{eq:imp appr}
U_{\Delta{t}} \simeq
\mathrm{e}^{-\mathrm{i}V_{\text{dip}}(x)\Delta{t}} \ ,
\ee
where, in force of the impulse approximation, we dropped both the kinetic term
and the harmonic oscillator potential.
In addition, we expand the dipole potential around the centre $x_0$ of the
harmonic trap
\be \label{eq:lin appr}
V_{\text{dip}}(x) \simeq V_{\text{dip}}(x_0)+V'_{\text{dip}}(x_0)(x-x_0) \ ,
\ee
where the prime denotes the first spatial derivative.
The largest possible value of $|x-x_0|$ is fixed by the extension of the
relative motion wavefunction $\psi$. This can be quantified by the average
position $\langle{x}\rangle=\int{x}|\psi|^2\,\text{d}x$ and the mean square
length
$\langle{\Delta{x}}\rangle=\sqrt{\langle{x^2}\rangle-\langle{x}\rangle^2}$, so
that the validity of the linear approximation (\ref{eq:lin appr})
requires%
\footnote{One could improve this approximation by linearizing
$V_{\text{dip}}(x)$ around $\langle{x}\rangle$ instead of $x_0$. This, however,
would make the calculation dependent on the initial motional state. A proper
discussion of the corresponding procedure could be done within the optimal
control technique.}
$|x-x_0|\lesssim|\langle{x}\rangle-x_0|+\langle{\Delta{x}}\rangle\ll{x_0}$.
A full description of the consequences of the above conditions in terms of
experimental parameters is given in Subsec.~\ref{subsec:approximations}.

With the previous approximations the time evolution operator acquires the
elementary form
\be
U_{\Delta{t}} \simeq \mathrm{e}^{-\mathrm{i}u\Delta{t}+\mathrm{i}\Delta{p}(x-x_0)} \ ,
\ee
where $u=V_{\text{dip}}(x_0)$ is the zero-energy shift and the momentum
transfer (`kick') $\Delta{p}$ is simply given by the classical expression
$\Delta{p}=-V'(x_0)\Delta{t}$.

\subsection{The dynamical phase}

The `ideal' picture described in Section~\ref{sec:scheme} can now be joined
with the actual form of the kick dynamics.
The phase space depicted in Fig.~\ref{fig:triangle} will clearly correspond to
the relative motion harmonic oscillator with frequency $\omega$ (which scales
to $1$ in dimensionless units),
whereas the kicks will be provided by the dipole potential.
As previously discussed, in first approximation one has
$p_1=|V_+'(x_0)\Delta{t}_1|$ and
$p_2=|V_-'(x_0)\Delta{t}_2|$, i.e., the momentum transfers
$p_{1,2}$ require a finite (although very short, as far as $V_{\mathrm{dip}}$
is large) time $\Delta{t}_{1,2}$. In addition, since $p_2=2p_1\cos\theta$ and%
\footnote{Here we assume that all the kicks are obtained by coupling the atoms
to the same Rydberg state, choosing the simplest experimental situation where
only one laser frequency is required. In principle, nothing prevents from using
more complicated schemes which add further degrees of freedom to the protocol.}
$V_-'(x_0)=-2V_+'(x_0)$, one has
$\Delta{t}_2=\Delta{t}_1\cos\theta$.
The total phase $\phi$ gained by the system is the sum of the dynamical phase
$\phi_{\text{dyn}}=-[V_+(x_0)\Delta{t}_1+V_-(x_0)\Delta{t}_2+V_+(x_0)\Delta{t}_1]=
-2V_+(x_0)\Delta{t}_1(1-\cos\theta)$ and the geometrical phase
$\phi_{\text{geom}}$. Hence
\be
\phi =
-2\frac{\alpha_+^2}{x_0^3}\Delta{t}_1(1-\cos\theta)
-\frac{9}{2}\frac{\alpha_+^4}{x_0^8}\Delta{t}_1^2\sin(2\theta) \ ,
\ee
which can be solved to give
\begin{eqnarray}
&&\alpha_+^2\Delta{t}_1 = \label{eq:triangle sol} \\
&&\frac{2}{9}\frac{x_0^5}{\sin(2\theta)}\left[\cos\theta-1+
\sqrt{(\cos\theta-1)^2+\frac{9}{2}\frac{\sin(2\theta)}{x_0^2}|\phi|}\right]
\ . \nonumber
\end{eqnarray}
This allows to choose the proper combination of Rydberg quantum numbers and
interaction time in order to achieve the desired phase.
The full gate time required by the operation will be
$\tau=\Delta{t}_1(2+\cos\theta)+2\theta\sim2\theta$.
It is also interesting to consider the fast quantum gate limit $\theta\to0$
where one easily finds
$\alpha_+^2\Delta{t}_1\sim(x_0^4/3)\sqrt{|\phi|/\theta}$. This implies
$\phi_{\text{dyn}}\sim-(x_0/3)\theta^{3/2}\sqrt{|\phi|}$ and
$\phi_{\text{geom}}\sim{\phi}$, which means that in this limit the global phase
$\phi$ is completely given by the geometrical contribution.

Let us stress again that the phase $\phi$ will arise only if the atoms are
initially in the $|gg\rangle$ internal state, since otherwise the dipole
potential is never switched on.

\subsection{Validity of the employed approximations}
\label{subsec:approximations}

Our impulse approximation (\ref{eq:imp appr}) requires that, during the
interaction time $\Delta{t}$, the action of the relative motion oscillator
Hamiltonian $H_0=p^2/2+(x-x_0)^2/2$ (we keep on using harmonic oscillator
units) be negligible%
\footnote{In this analysis we assume for simplicity that $\omega$ and $\omega'$
are of the same order \cite{jaksch}.}.
The contributions of the different terms in the Hamiltonian can be quantified
at a given time by taking their expectation values on the corresponding state
$|\psi(t)\rangle$.
We therefore require
\be \label{eq:imp cond}
\langle{H_0}\rangle\ll|\langle{V_{\mathrm{dip}}}\rangle|
\ee
during the whole kick time $\Delta{t}$.
However, from the point of view of the gate dynamics, it is also necessary to
have
\be \label{eq:phi cond}
\langle{H_0}\rangle\Delta{t}\ll|\phi| \ ,
\ee
since otherwise we should include such a contribution in the calculation of the
dynamical phase $\phi_{\mathrm{dyn}}$.

Besides conditions (\ref{eq:imp cond}) and (\ref{eq:phi cond}), we also have to
consider the validity requirements of the linear approximation (\ref{eq:lin
appr}), i.e., the condition
\be \label{eq:lin cond}
|\langle{x}\rangle-x_0|+\langle{\Delta{x}}\rangle\ll{x_0} \ ,
\ee
necessary for the linearization of the dipole potential.
Clearly, the explicit form of all the previous conditions depends on the states
reached by the system during the dynamics of the kick scheme.

For example, if the system stays very close to the oscillator ground state
during the whole gate time, one can estimate the above expectation values by
using $|\psi(t)\rangle=|0\rangle$. In this case $\langle{H_0}\rangle=1/2$,
$\langle{x}\rangle=x_0$ and $\langle{\Delta{x}}\rangle=1/\sqrt2$. Linearization
then requires $x_0\gg1/\sqrt2$, while, by assuming the validity of the latter
condition, Eq.~(\ref{eq:imp cond}) reduces to $1/2\ll{V_+(x_0)}$, which can be
made explicit by evaluating $V_+(x_0)=\alpha_+^2/x_0^3$ with
Eq.~(\ref{eq:triangle sol}). Eq.~(\ref{eq:phi cond}) is simply
$\Delta{t}\ll2|\phi|$.

Since we want our protocol to be independent of the initial motional
state, it is interesting to find the conditions under which our approximations
are applicable to a generic state {\em not too far} from the ground state.
We therefore consider the case $|\psi(t)\rangle=|\alpha(t)\rangle$, where
$|\alpha(t)\rangle$ is a coherent state subject to the constraint
$|\alpha(t)|\leq{R}$,
${R}>0$.
Then, if the initial state of the system is the coherent state
$|\alpha(0)\rangle$, the boundary $R$ can be estimated within the ideal kick
scheme, where $|\alpha(t)|\leq|\alpha(0)|+\tilde{p}_1$.
Concerning the expectation values we have
$\langle{H_0}\rangle=|\alpha|^2+1/2\leq{R}^2+1/2$,
$|\langle{x-x_0}\rangle|=\sqrt2\,|\mathrm{Re}\,\alpha|\leq\sqrt2{R}$, and still
$\langle{\Delta{x}}\rangle=1/\sqrt2$.
If $x_0\gg\sqrt2{R}+1/\sqrt2$ then the use of the linear approximation is
legitimate and Eq.~(\ref{eq:imp cond}) becomes ${R}^2+1/2\ll{V_+(x_0)}$, while
Eq.~(\ref{eq:phi cond}) is $({R}^2+1/2)\Delta{t}\ll2|\phi|$.

We explicitly evaluate these conditions for the most interesting case
$\alpha(0)=0$, where initially the system is in the oscillator ground state and
hence $R=\tilde{p}_1=(3/\sqrt2)V_+(x_0)\Delta{t}_1/x_0$. Qualitatively, fast
gate performances require small values of $\theta$, while the linear
approximation (\ref{eq:lin appr}) needs large values of $x_0$. However large
distances disfavour the impulse approximation (\ref{eq:imp appr}), which is
valid for strong dipole interactions. High Rydberg quantum numbers are then
necessary to enhance the dipole moments. In order to study the interplay among
the different parameters it is useful to consider Eq.~(\ref{eq:triangle sol})
in the limits $\theta\to0$ and $x_0\to\infty$.
Then, it turns out that in all the cases of practical interest condition
(\ref{eq:imp cond}) is already implied by Eqs.~(\ref{eq:phi cond}) and
(\ref{eq:lin cond}), so that one is left with only two conditions. In addition,
the most stringent form of these requirements takes place in the $\theta\to0$ limit,
where $V_+(x_0)\Delta{t}_1\sim(x_0/3)\sqrt{|\phi|/\theta}$. In conclusion, the
linearization condition (\ref{eq:lin cond}) and Eq.~(\ref{eq:phi cond})
respectively reduce to
\begin{eqnarray}
&&x_0\gg\sqrt{|\phi|/\theta} \ , \label{eq:x_0 cond} \\
&&\Delta{t}_1\ll2\theta \ , \label{eq:Dt1 cond}
\end{eqnarray}
valid in the fast-gate limit $\theta\to0$.

The conditions (\ref{eq:x_0 cond}-\ref{eq:Dt1 cond}) are actually rather
demanding in terms of experimental parameters, even though certainly
achievable. Assuming Eqs.~(\ref{eq:x_0 cond}) and (\ref{eq:Dt1 cond}) to be
well satisfied by a factor of ten difference between the two terms of each
inequality, so that $x_0=10\sqrt{|\phi|/\theta}$ and $\Delta{t}_1=\theta/5$,
and recalling the $\theta\to0$ result
$\alpha_+^2\sim(x_0^4/3\Delta{t}_1)\sqrt{|\phi|/\theta}$, one gets
$\alpha_+^2\sim(10^5/6\theta)(|\phi|/\theta)^{5/2}$.
Unfortunately, $\alpha_+^2$ cannot be arbitrarily large.
For example, for the case of $^{87}$Rb atoms, where $\mu=87/2$~amu, with a
harmonic frequency $\omega=2\pi\times100$~KHz, so that
$(\mu/m_e)/(a_{\text{ho}}/a_0)=86.8$, by choosing the Rydberg quantum numbers
$n_{1,2}=q_{1,2}+1=50$ one has $\alpha_+^2=1.2\times10^9$.
Then, for $|\phi|=\pi$ (equivalent to the C-NOT gate), the shortest gate-time
obtainable within this rough upper bound for $\alpha_+^2$ corresponds to
$\theta=[(10\sqrt{|\phi|})^5/6\alpha_+^2]^{2/7}=0.09$, i.e.,
$\tau\sim0.03T_{\mathrm{ho}}$, where $T_{\mathrm{ho}}=2\pi$ is the
harmonic oscillator period.
Notice also that, while the condition corresponding to Eq.~(\ref{eq:Dt1 cond}),
$\Delta{t}_1=\theta/5$, can be easily satisfied experimentally,
Eq.~(\ref{eq:x_0 cond}) for such a small value of $\theta$ imposes a very large
separation between the harmonic wells (in harmonic oscillator units), namely
$x_0=10\sqrt{|\phi|/\theta}=58$, significantly larger than the value
corresponding to the typical distance between neighbouring sites in currently
employed optical lattices (for example, for a laser wavelength of $800\,$nm,
the above parameters yield $x_0=8.3$).
However, this requirement can be clearly relaxed by increasing the gate time
(see the next section).
Besides, there is the possibility of increasing $x_0$ independently
of the lattice spacing, e.g., by using superlattices or the methods discussed
in Ref.~\cite{lens}.

In the final part of this section, we briefly discuss the validity of the
one-dimensional approximation (see also \cite{jaksch}). We assume a harmonic
trapping also in the radial direction, with a frequency
$\omega_\perp$ and restore the dimensional units. The radial
degrees of freedom could then be neglected if the level spacing
$\hbar\omega_\perp$ were much larger than all the other energy
scales which come into play. Clearly, this is difficult to
accomplish as far as one also needs large values of the frequency
$\omega$ along the $x$ direction, necessary to increase the ratio
$x_0/a_{\mathrm{ho}}\propto\sqrt\omega$. However, provided each
atom is sitting in the ground state of the corresponding harmonic
well, the dimensionality does not play any role. Indeed, assuming
the exact dipole polarization, the momentum transfer takes place
only in the $x$ direction and the gate dynamics is decoupled from
the radial degrees of freedom, irrespectively of the value of
$\omega_\perp$. In the case where the initial motional state
significantly differs from the ground state, one instead has to
compare $\hbar\omega_\perp$ with $\hbar\omega$ and with the
component of the dipole potential affecting the radial motion,
which is however smaller and smaller as the inter-well distance
increases. Consequently, the validity condition of the linear
approximation also favours the quenching of 3D effects, even
though a quantitative analysis of this approximation would require
a full description of the 3D dynamics, which is beyond the scope
of this paper.

\section{Fidelity analysis}
\label{sec:errors}

We are now interested in considering possible error sources. The effects we
take into account in the following are (i) contributions not included in the
impulse and kick approximations, which are important when the validity
conditions discussed in the previous section are not well satisfied, and (ii)
anharmonic terms in the trapping potential. Notice that, since as far as our
approximations are valid the model is independent of the initial motional state,
finite `temperature' (in a sense specified later) effects are actually included
in the first point.

Note that in addition, as discussed in \cite{jaksch}, minor error sources that are
completely neglected here are the spontaneous emission and the black body radiation
for the Rydberg state.

In our scheme atoms in different wells have a negligible overlap and can be
considered distinguishable. In this case the gate fidelity can be defined as
\cite{schumacher}
\be
F = \min_{\chi} \{\textrm{Tr}_{\textrm{ext}}[\langle\tilde\chi|
\mathcal{U}_\tau(|\chi\rangle\langle\chi|\otimes\rho_0)\mathcal{U}_\tau^\dagger
|\tilde\chi\rangle]\} \ ,
\label{eq:fid}
\ee
where $\mathcal{U}_\tau$ is the unitary operator governing the evolution of the
full system during a single gate operation,
$|\chi\rangle=\sum_{\alpha\beta}c_{\alpha\beta}|\alpha\beta\rangle$ is an
arbitrary internal state of the two atoms with the normalization
$\sum_{\alpha\beta}|c_{\alpha\beta}|^2=1$, $|\tilde\chi\rangle$ is the state
given by the application of the ideal gate (\ref{eq:truth}) to $|\chi\rangle$,
and $\rho_0$ is the initial density operator relative to the external degrees
of freedom, over which the trace is taken.
In our case, the external states are simply given by the motional states.
Furthermore, since the centre of mass motion is unaffected by the
gate, we can restrict the analysis to the relative motion%
\footnote{This is not generally true, e.g., if the atoms were in the internal
state $|er\rangle$ for a time comparable with the oscillator time scales.}.

\subsection{Fidelity for pure motional states}
\label{subsec:S=0}

Let us consider the particular case $\rho_0=|k\rangle\langle k|$, where
$|k\rangle$ are the eigenstates of the relative motion harmonic
oscillator.
This corresponds to assuming zero entropy, the system being in a pure state.

We can write the action of $\mathcal{U}_\tau$ as
\begin{eqnarray}
\label{eq:U}
\mathcal{U}_\tau(|\chi\rangle\otimes|k\rangle) & = &
{\sum_{\alpha\beta}}'c_{\alpha\beta}|\alpha\beta\rangle
\otimes|k\rangle\mathrm{e}^{-i(k+1/2)\tau}+ \\
&&+c_{gg}|gg\rangle\otimes\sum_{k'}\alpha_{kk'}|k'\rangle\mathrm{e}^{-\mathrm{i}(k'+1/2)\tau}
\ , \nonumber
\end{eqnarray}
where the primed summation does not contain the term
$\alpha\beta=gg$, and $\sum_{k'}|\alpha_{kk'}|^2=1$.
The explicit calculation then gives
\begin{eqnarray}
\nonumber F_k & = &
\min_{|c_{gg}|} \{1+2(1-|c_{gg}|^2)|c_{gg}|^2
[\textrm{Re}(\mathrm{e}^{\mathrm{i}\phi}\alpha_{kk}^*)-1]\} = \\
&=&\frac{1+|\alpha_{kk}|\cos(\phi-\phi_{kk})}{2} \ , \label{eq:F_k}
\end{eqnarray}
where $\phi_{kk}=\arg\alpha_{kk}$ and $\mathrm{e}^{\mathrm{i}\phi}\alpha_{kk}^*=
\langle{\psi_{\mathrm{num}}(k)|\psi_{\mathrm{id}}(k)}\rangle$ is the overlap
between $|\psi_{\mathrm{id}}(k)\rangle=\mathrm{e}^{\mathrm{i}\phi}|k\rangle$ and the final
motional state obtained by applying the full gate dynamics to the state
$|gg\rangle\otimes|k\rangle$.

We compute the zero-entropy fidelity in the good initialization case $k=0$.
In the ideal case where the impulse approximation is valid one has
$\alpha_{00}=\mathrm{e}^{\mathrm{i}\phi_{00}}=\mathrm{e}^{\mathrm{i}\phi}$ and $\alpha_{0k}=0$ for
$k\geq1$.

\begin{figure}
\includegraphics[angle=-90,width=8cm]{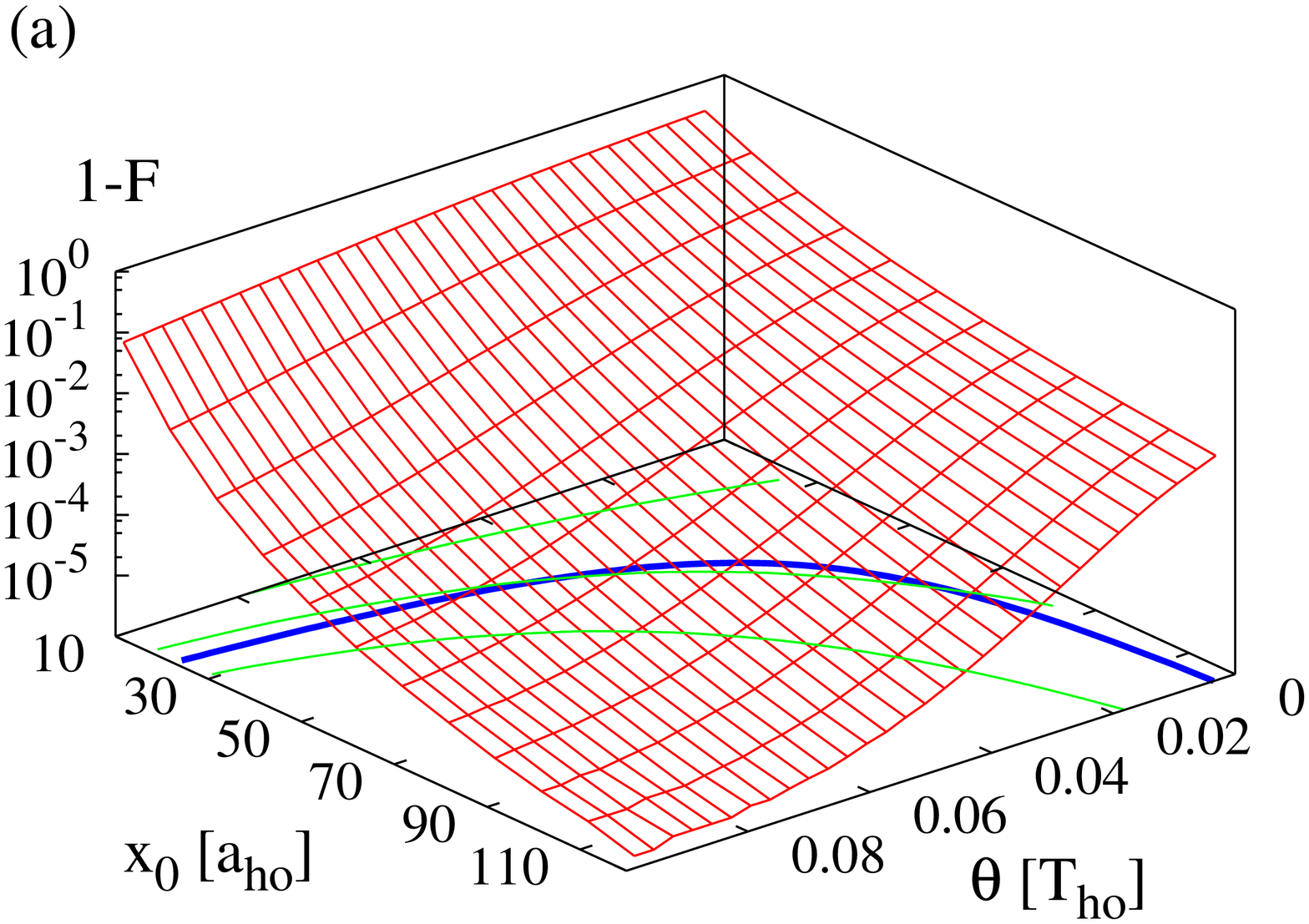}
\includegraphics[angle=-90,width=8cm]{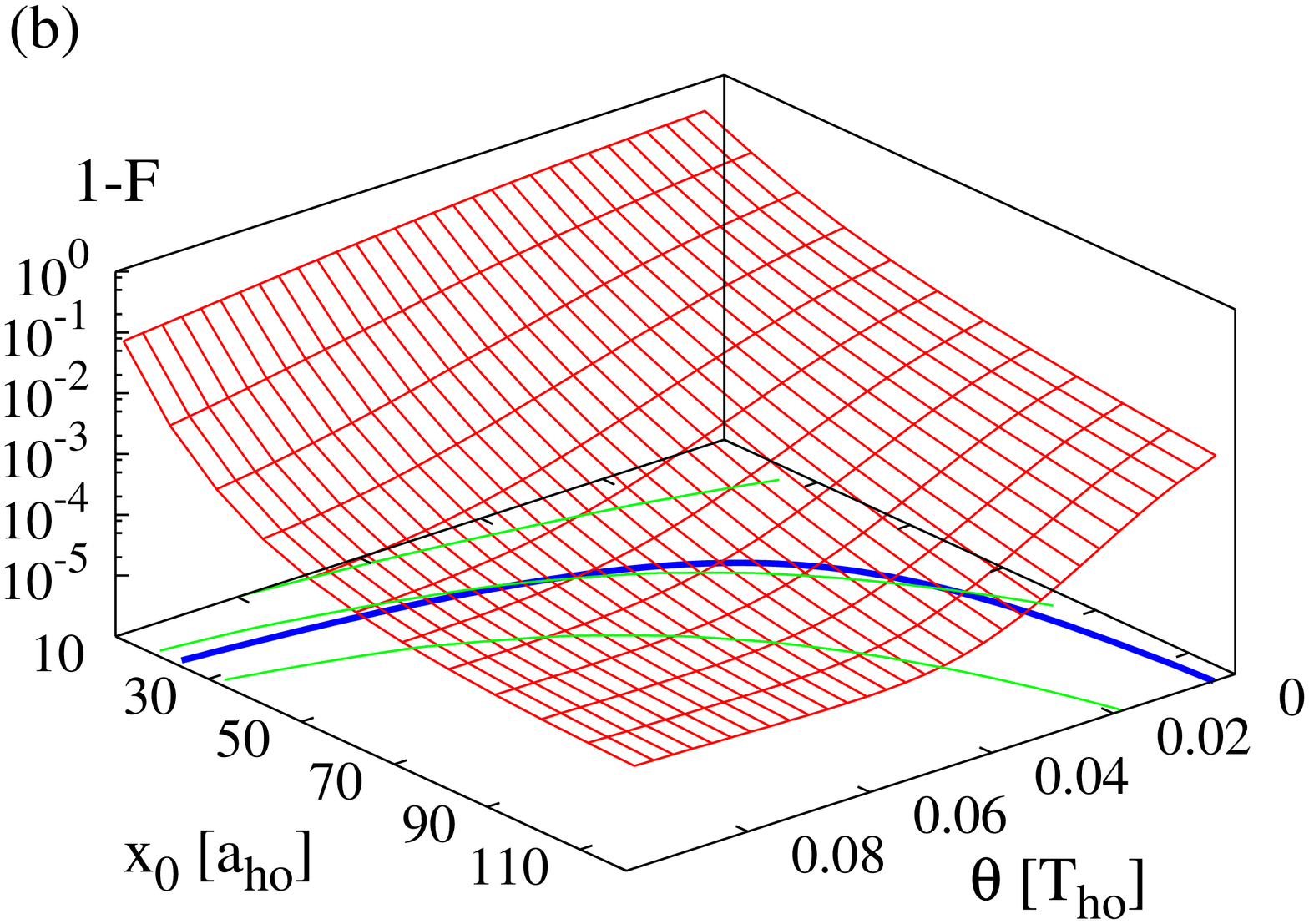}
\caption{\label{fig:fid}%
Upper panel (a): fidelity
$F=(1+\textrm{Re}\langle{\psi_{\mathrm{id}}|\psi_{\mathrm{num}}}\rangle)/2$
as a function of the time $\theta$ and the distance $x_0$ (harmonic oscillator
units), for $|\phi|=\pi$ and $\Delta{t}_1=10^{-4}\times2\pi$.
The state $|\psi_{\mathrm{num}}\rangle$ is obtained by calculating
numerically the full gate-dynamics (in the absence of anharmonic perturbations)
for the state $|gg\rangle\otimes|0\rangle$, while the ideal state is simply
$|\psi_{\mathrm{id}}\rangle=\mathrm{e}^{i\phi}|0\rangle$.
The contour lines drawn in the lower plane correspond to the fidelity values
$0.9$,$0.99$,$0.999$, while the thick line plots the relation
$x_0=10\sqrt{\pi/\theta}$.
Lower panel (b): same as (a) but in the presence of an anharmonic quartic term
with $\lambda=10^{-2}$.}
\end{figure}

In Fig.~\ref{fig:fid}(a) we show the gate fidelity $F$ when the protocol
is applied to the motional ground state ($k=0$), as a function of the two most
important parameters $\theta$, $x_0$. This allows a direct
comparison with the conditions discussed in
Subsec.~\ref{subsec:approximations}. The data are obtained by
solving numerically the time dependent Schr\"odinger equation
corresponding to the full kick dynamics applied to the oscillator
ground state, for the case of purely harmonic trapping. The
behaviour corresponding to Eq.~(\ref{eq:x_0 cond}) is also
plotted, showing a qualitative agreement with the curves at
constant fidelity. Indeed, for the considered parameters,
Eq.~(\ref{eq:Dt1 cond}) is always very well satisfied and the
terms not included in the linear approximation are the main reason
of discrepancy with respect to the ideal scheme.

In Fig.~\ref{fig:fid}(b) we repeat the same analysis by adding to $H_0$ an
anharmonic term $-\lambda(x-x_0)^4/2$, corresponding to the first correction
one would have for the square sinus potential of an optical lattice%
\footnote{Actually, the square sinus potential enters the single atom
Hamiltonians. In this case it is not even possible to exactly separate the
centre of mass and relative motion Hamiltonians. Such decoupling is indeed a
peculiar property of the harmonic potential. However, the quartic correction
can be used as a crude test of anharmonic effects.}.
Only a very slight worsening of the fidelity is observed for the realistic
value $\lambda=10^{-2}$, showing that in practice the validity of the linear
approximation is the crucial condition.

\subsection{Finite temperature and entropy effects}
\label{subsec:entropy}

In view of a realistic quantum computing scenario it is important to
investigate the dependence of the gate performances on temperature.
Indeed, in the course of the computational process the system is expected to
undergo a certain heating -- for example due to moving the atomic qubits around
by shifting the optical lattice -- which could in principle be limited by a
periodic re-cooling of the qubits \cite{daley}.

On the other side, atoms in optical lattices can be considered practically
isolated from the environment, so that in the
experimental setup considered here atoms do not easily thermalize.
An alternative interesting quantity to study is hence given by entropy, whose
value can be extracted from the density matrix of the system.

\begin{figure}[t]
\begin{center}
\includegraphics[width=8cm]{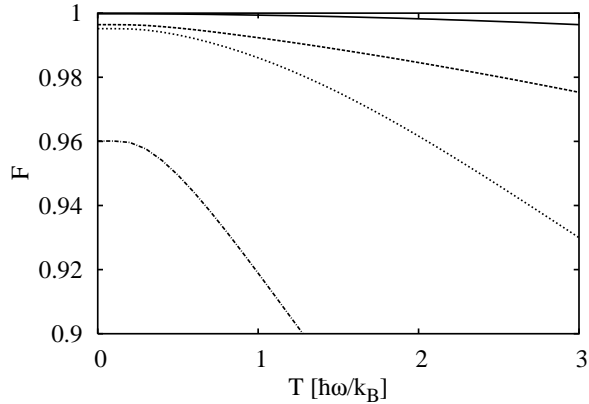}
\caption{Temperature dependence of the fidelity
($|\phi|=\pi$, $\Delta{t}_1=10^{-4}T_{\mathrm{ho}}$, $\lambda=0$).
Curves from top to bottom correspond to
$\theta=0.05T_{\mathrm{ho}}$, $x_0=20$ (solid line),
$\theta=0.05T_{\mathrm{ho}}$, $x_0=40$ (dashed line),
$\theta=0.1T_{\mathrm{ho}}$, $x_0=20$ (dotted line),
$\theta=0.1T_{\mathrm{ho}}$, $x_0=40$ (dot-dashed line).}
\label{fig:temp}
\end{center}
\end{figure}

Let us first discuss the fidelity of the gate in the case where the density
matrix of the relative motion is given by the canonical thermal
distribution
\be \label{eq:rho_T}
\rho_{\text{rel}}(T) = 2\sinh\frac{E_0}{k_BT}
\sum_{k}\mathrm{e}^{-E_k/k_BT}|k\rangle\langle{k}| \ ,
\ee
where $T$ is the temperature, $k_B$ is the Boltzmann constant, and $E_k$ are
the oscillator energies.
The density operator satisfies the normalization
$\textrm{Tr}[\rho_{\text{rel}}(T)]=1$.

By substituting $\rho_0=\rho_{\mathrm{rel}}(T)$ into Eq.~(\ref{eq:fid}),
the fidelity of the finite temperature motional state is found to be
\be
F = 2\sinh\frac{E_0}{k_BT}\sum_k\mathrm{e}^{-E_k/k_BT}F_k \ ,
\ee
where $F_k$ is given in Eq.~(\ref{eq:F_k}).
The resulting behaviour for reasonable values of the (purely harmonic,
$\lambda=0$) trap parameters is reported in Fig.~\ref{fig:temp}. Notice that
current experimental techniques allow to cool the system much below the
highest temperature considered in the figure (which is three times the
temperature defined by the oscillator energy). In the numerical calculation,
for each considered temperature the density matrix is constructed by including
in Eq.~(\ref{eq:rho_T}) all the eigenstates up to at least $E_k=5k_BT$.
It is worth mentioning that the presence of anharmonicities in the trapping
potential can worsen the large temperature behaviour of the curves shown in
Fig.~\ref{fig:temp}. Indeed, higher oscillator states, having a larger spatial
extension, are more sensitive to deviations from the quadratic confinement.
However, for a quartic term with $\lambda=10^{-2}$ (see previous subsection)
and the parameters of Fig.~\ref{fig:temp} this becomes important only for
$k_BT\gtrsim3$ (corresponding to the mean occupation number $\bar{k}\sim2.5$).

As anticipated above, it also useful to look at the entropy increase
caused by the gate evolution.
In fact, even if the very first initialization was perfectly done, i.e.,
$\rho_0=|0\rangle\langle0|$, after a certain number of operations one expects
the system to be in a superposition of excited motional states.
As shown by Eq.~(\ref{eq:U}), already after a single 2-qubit gate the
system is left in a non-trivial, entangled state of the internal and external
degrees of freedom, which will spoil to some extent a successive application of
the phase gate.
Clearly, for the gate to make sense this must be a small effect, which is
guaranteed by fidelity values close to $1$.
Qualitatively, it is also interesting to see which is the fidelity for the
finite entropy states arising from repeated executions of the gate operation.
Indeed, although in practice computation would not involve always the same two
qubits, this gives an idea of the propagation of the gate error.

\begin{figure}[t]
\begin{center}
\includegraphics[angle=-90,width=8cm]{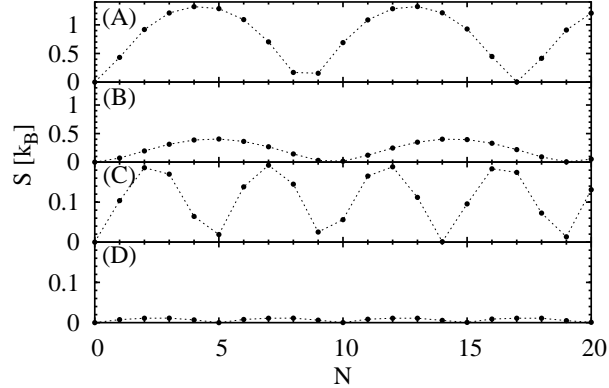}
\caption{Entropy $S$ (in units of $k_B$) as a function of the number of cycles
$N$ for $|\phi|=\pi$ and $\Delta{t}_1=10^{-4}T_{\mathrm{ho}}$.
The considered trapping potential is the purely harmonic ($\lambda=0$) and the
remaining parameters are
(A) $\theta=0.05T_{\mathrm{ho}}$, $x_0=20$,
(B) $\theta=0.05T_{\mathrm{ho}}$, $x_0=40$,
(C) $\theta=0.1T_{\mathrm{ho}}$, $x_0=20$,
(D) $\theta=0.1T_{\mathrm{ho}}$, $x_0=40$.}
\label{fig:entr}
\end{center}
\end{figure}

Concretely, if we start with the system in the product state
$|\Psi\rangle=|\chi\rangle \otimes|0\rangle$, with the same notation as in the
previous subsection, after $N$ gate operations $|\Psi\rangle$ is transformed
into the $N$-step state $|\Psi(N\tau)\rangle=
c_{gg}|gg\rangle\otimes\sum_k\alpha_{0k}(N\tau)|k\rangle\mathrm{e}^{-\mathrm{i}(k+1/2)N\tau}+
(c_{ge}|ge\rangle+c_{eg}|eg\rangle+c_{ee}|ee\rangle)
\otimes|0\rangle\mathrm{e}^{-\mathrm{i}N\tau/2}$.
For simplicity, we consider the case $c_{gg}=1$ and construct a relative motion
density matrix with the $N$-step coefficients $\alpha_k(N\tau)$
\be \label{eq:rho_N}
\rho_{\text{rel}}(N) = \sum_{k}|\alpha_{0k}(N\tau)|^2|k\rangle\langle{k}| \ .
\ee
The entropy is
$S=-k_B\text{Tr}[\rho_{\text{rel}}(N)\ln\rho_{\text{rel}}(N)]$, where the trace
can be calculated summing over all the eigenvalues, already explicit in the
diagonal form of Eq.~(\ref{eq:rho_N}).

The entropy behaviour as a function of the number of gate cycles is plotted for
reasonable parameters in Fig.~\ref{fig:entr}.
In the numerical calculation the entropy is extracted by projecting the
time-evolved state onto the harmonic oscillator eigenstates $|n\rangle$. All
the eigenstates up to $n=25$ are included, much more than required for the
considered parameters, ensuring that the normalization is always preserved.
The maximum (numerical) entropy is $S/k_B=\ln n\sim 3.2$.
Note the periodic behaviour of the entropy: its value does not increase
arbitrarily. Indeed, after a given number of cycles one has a revival of the
initial state (i.e., only the relative motion ground state is occupied). This
is true only for reasonable parameters: for parameters which yield a bad gate
fidelity one does not have this revival and the entropy increases
monotonically.

\begin{table}[t]
\centering
\begin{tabular}{|c||c|c|}
\hline
        &   $\theta=0.05\,T_{\mathrm{ho}}$  &   $\theta=0.1\,T_{\mathrm{ho}}$ \\
\hline
\hline
$x_0=20$    &
    $\begin{array}{rcl}
    F_0\!&\!=\!&\!0.9601 \\
    F(N=4)\!&\!=\!&\!0.8889 \\
    S(N=4)\!&\!=\!&\!1.32
    \end{array}$    &
        $\begin{array}{rcl}
        F_0\!&\!=\!&\!0.9952 \\
        F(N=2)\!&\!=\!&\!0.9946 \\
        S(N=2)\!&\!=\!&\!0.19
        \end{array}$ \\
\hline
$x_0=40$    &
    $\begin{array}{rcl}
    F_0\!&\!=\!&\!0.9964 \\
    F(N=4)\!&\!=\!&\!0.9956 \\
    S(N=4)\!&\!=\!&\!0.39
    \end{array}$    &
        $\begin{array}{rcl}
        F_0\!&\!=\!&\!0.9998 \\
        F(N=2)\!&\!=\!&\!0.9998 \\
        S(N=4)\!&\!=\!&\!0.01
        \end{array}$ \\
\hline
\end{tabular}
\caption{\label{tab:1}%
Fidelity of a single gate execution at different entropy values for the same
parameters as in Fig.~\ref{fig:entr}.
$F_0$ is the fidelity calculated for the matrix
$|{\chi}\rangle\langle{\chi}|\otimes|{0}\rangle\langle{0}|$ (zero entropy),
while $F(N)$ for the matrix
$|{\chi}\rangle\langle{\chi}|\otimes\rho_{\mathrm{rel}}(N)$.
The entropy $S(N)$ is the same shown in Fig.~\ref{fig:entr}.}
\end{table}

As an estimate of the fidelity due to a non-zero entropy motional state we use
Eq.~(\ref{eq:fid}) with
$\rho_0=|{\chi}\rangle\langle{\chi}|\otimes\rho_{\mathrm{rel}}(N)$.
The calculation now gives
\be
F = \sum_k|\alpha_{0k}(N\tau)|^2F_k \ ,
\ee
and the corresponding results for some relevant values of $N$ are reported in
Table~\ref{tab:1}, showing that the fidelity is not affected very much by the
considered process.
Indeed, for the used parameters the coefficients $|\alpha_{0k}(N\tau)|^2$
entering Eq.~(\ref{eq:rho_N}) decay very rapidly by increasing $k$.
In other terms, although not directly comparable to the canonical distribution
(\ref{eq:rho_T}), the population of the first few states shows a behaviour
corresponding to $k_BT\ll1$ (in harmonic oscillator units).
As a consequence, we also notice that the small anharmonicity
($\lambda=10^{-2}$) previously considered does not alter significantly the
results presented in Table~\ref{tab:1} for the purely harmonic case.

\section{Conclusions and outlook}
\label{sec:conclusions}

In this paper we aimed at applying the method of coherent control
of two-body dynamics, already proposed in \cite{garcia} as a means
for obtaining fast, high-fidelity quantum gates with ions, to the
case of Rydberg-excited atoms. We found that the method works well
in a realistic situation with a simple three-pulse sequence that
allows for obtaining a two-qubit control-phase gate with a
fidelity bigger than 99.9\% in a fraction of the trap period for
presently achievable experimental parameters.
While the ideal scheme at the basis of the gate is completely independent
of the initial motional state of the atoms, the approximations employed to
treat the dipole-dipole interaction
introduce deviations which
worsen the gate performance for motional states different from the
ground state.
We then analyze the effect of errors in the preparation of the motional state
in terms of temperature and entropy. For parameters reasonably close to the
ones optimized for high fidelity, the temperature of the initial state is found
to significantly affect the gate quality only for relatively large values,
comparable with the oscillator energy of the trapping potential.
In addition, the entropy of the evolved state obtained from repeated
applications of the gate to the same two qubits does not increase indefinitely
with the number of operations, but it rather oscillates in a quite regular
fashion, exhibiting full ground-state revivals after a few gate iterations.
This is likely to depend on symmetries in the underlying evolution, evidenced
by the trivial iteration of the gate on the same two qubits.
Even though the latter analysis does not convey much information on a real
quantum algorithm, which typically involves many gates between different
qubits, it would be interesting to study the possibility that similar
recurrences emerge over more complex gate sequences. In this case, indeed,
optimization could be done not only at the level of a single quantum gate, but
also on a broader time scale involving ``blocks'' of subsequent gates, in the
sense of tailoring each gate's parameters in order to minimize the system
entropy after a given gate sequence.
A deeper understanding of this aspect is still missing, as it goes beyond the
scope of the present work, and shall be the subject of future investigations.

Let us finally comment on a possible alternative implementation of
the phase-space displacements which allow to acquire the
geometrical phase. Indeed, the momentum shifts obtained from the
dipole kicks could in principle be substituted by spatial shifts,
straightforwardly realizable by displacing the centre of the
harmonic oscillator potential, provided that one manages to
properly design the spin dependency of the trapping potential
\cite{bloch}.

\begin{ack}

This work was partially supported by the European Commission under
Contracts No. IST-2001-38863 (ACQP), FP6-013501-OLAQUI,
FP6-015714-SCALA, and by the National Science Foundation through a
grant for the Institute for Theoretical Atomic, Molecular and
Optical Physics at Harvard University and Smithsonian
Astrophysical Observatory. Work at the University of Innsbruck is
supported by the Austrian National Science Foundation and the EU
projects.

\end{ack}

\end{document}